\documentstyle[twoside,epsfig,psfig]{article} 
 
\input ibvs2.sty 
 
\begin{document} 
\IBVShead{6176}{6 July 2016} 
 
\IBVStitle{Historical light-curve and the 2016 outburst of}
\vskip -0.3cm
\IBVStitle{the symbiotic star StHa 169}
\IBVStitle{} 
  
\IBVSauth{Ulisse Munari$^1$, Mauro Graziani$^2$, and Rajka Jurdana-{\v S}epi{\'c}$^3$} 

\IBVSinst{INAF Osservatorio Astronomico di Padova, Sede di Asiago, I-36032 Asiago (VI), Italy} 
\IBVSinst{ANS Collaboration, c/o Astronomical Observatory, 36012 Asiago (VI), Italy} 
\IBVSinst{Physics Department, University of Rijeka, Radmile Matej\v{c}i\'{c}, 51000, Rijeka, Croatia}

\IBVStyp{Symbiotic star} 
\IBVSkey{photometry,spectroscopy} 
\IBVSabs{Accurate BVRI optical photometry of StHa 169 has been collected
during the period 2005-2016.  It reveals two outbursts, in 2009 and 2016,
both peaking at B=13.6.  The mean brightness in quiescence was B=15.29.  In
response to the 2016 outburst, the integrated absolute flux of the emission
lines increased by a factor of 7 for Balmer, 9 for HeI lines, and 4.5 for
HeII 4686, the Balmer continuum turned into strong emission, and [NeV] 3526
and OVI Raman 6825 Ang emissions disappeared.  We have reconstructed the
1897-1951 B-band lightcurve of StHa 169 from Harvard historical plates.  On
top of a slow decline from a large outburst, ending in 1916, several
brightenings are visible.  Two fast-evolving outbursts, separated by 510
days, occurred in 1934 and 1935, peaking at B=13.7.  The brightness in
quiescence was the same as we measured in 2005-2016.}

\begintext 

StH$\alpha$ 169 (J2000 $\alpha$=19$^h$ 49$^m$ 57$^s$.59,
$\delta$=+46$^\circ$ 15' 20''.6) was discovered by Stephenson (1986) during
an objective prism search for emission line objects away from the Galactic
plane.  No information was logged on the type of spectrum or intensity of
H$\alpha$ emission, only its magnitude was recorded as $m_V$$>$13.5.  The
symbiotic nature of StH$\alpha$ 169 was recorgnized by Downes \& Keyes
(1988) in the course of a spectroscopic survey of Stephenson (1986) objects. 
Their spectrum shows Balmer, HeI and HeII 4686 emission lines superimposed
to the absorption spectrum of an M2 giant, similarly to what recently
reported by Li et al.  (2015) from a LAMOST survey spectrum.  The
classification by Downes \& Keyes (1988) prompted the inclusion of the star
in the catalog of symbiotic stars compiled by Belczy{\'n}ski et al.  (2000). 
Henden \& Munari (2008) reported UBVRI photometry at three epochs in 2001,
their mean values being $V$=13.68, $U-B$=+0.952, $B-V$=+1.64, $V-R_{\rm
C}$=+1.04, and $V-I_{\rm C}$=+2.14.  Pigulski et al.  (2009) obtained
$V$,$I_{\rm C}$ photometry of StH$\alpha$ 169 from June 2006 to Jan 2008,
with mean values $V$=13.44 and $V-I_{\rm C}$=+2.03.  Their short focal
length did not resolved StH$\alpha$ 169 from a nearby field star, 10 arcsec
to the East, for which Henden \& Munari (2006) give $V$=16.606,
$U-B$=+0.946, $B-V$=+1.208, $V-R_{\rm C}$=+0.745, and $V-I_{\rm C}$=+1.519. 
Correcting the Pigulski et al.  (2009) photometry of StH$\alpha$ 169 for the
contribution of this nearby field star provides $V$=13.50 and $V-I_{\rm
C}$=+2.05, close to the 2001 values measured by Henden \& Munari (2008). 
The star is situated within the field of view of the planet-hunter {\it
Kepler} space mission.  According to Ramsay et al (2014) the Kepler
unfiltered, white-light 2009-2013 data shows a quasi-periodic behaviour with
a mean period of 34 days and an amplitude of a few per cent superimposed on
a stable mean brightness, consistent with a low amplitude variability
intrinsic to the cool giant.  To a good approximation, this is all what is
known about StH$\alpha$ 169, which can thus be appropriately labelled as one
of the poorest studied symbiotic stars.  We are continously monitoring
StH$\alpha$ 169 since 2005, both photometrically and spectroscopically, and
have recently reported on it entering an outburst state during 2016 (Munari
\& Graziani 2016).

$B$$V$$R_{\rm C}$$I_{\rm C}$ optical photometry of StH$\alpha$ 169 is
regularly obtained with ANS Collaboration telescope N.  73, a 0.30-m f/10
Meade LX200 telescope located in Alfonsine (Ravenna, Italy).  It is equipped
with $U$$B$$V$$R_{\rm C}$$I_{\rm C}$ Astrodon filters.  The CCD is an Finger
Lake Instruments MAXCAM CM9-1E 512$\times$512 array, 20 $\mu$m pixels
$\equiv$ 1.37$^{\prime\prime}$/pix, with a field of view of
11$^\prime$$\times$11$^\prime$.  Image quality and plate scale allow full
separation of the variable from the nearby field star above described, to
the point that no difference is found between the results obtained with
aperture photometry or PSF-fitting.  The local photometric sequence,
calibrated by Henden \& Munari (2006) against Landolt equatorial standards,
was used throughout the whole observing campaign, ensuing a high consistency of
the data.  Our $B$$V$$R_{\rm C}$$I_{\rm C}$ photometry of StH$\alpha$ 169 is
given in Table~1 (available electronic only), where the quoted uncertainties
are the total error budget, which quadratically combines the measurement
error on the variable with the error associated to the transformation from
the local to the standard photometric system (as defined by the photometric
comparison sequence).  A detailed description of ANS Collaboration
telescopes operation and data reduction is provided by Munari et al.  (2012)
and Munari \& Moretti (2012).

\IBVSfig{9.5cm}{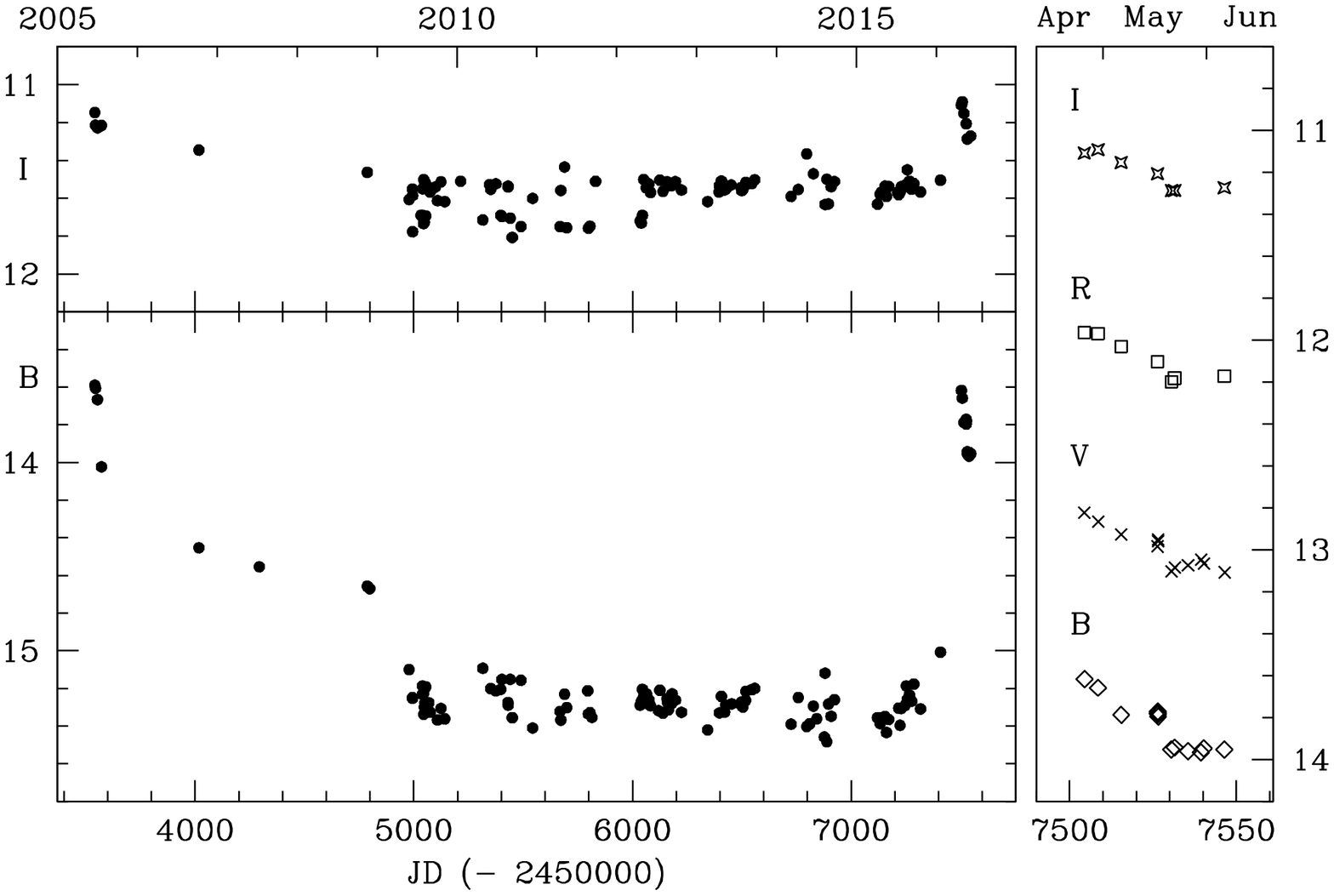}{{\it Left :} overall 2005-2016 light-curves in
the $B$ and $I_{\rm C}$ bands of StH$\alpha$ 169.  {\it Right :} a zoom in
all four $B$$V$$R_{\rm C}$$I_{\rm C}$ bands on the 2016 outburst.}

The 2005-2016 lightcurve of StH$\alpha$ 169 based on the data in Table~1 is
presented in Figure~1.  During 2005-2009 the variable appears declining from
a large amplitue outburst ($\Delta B$$\sim$2 mag), which maximum could have
occurred at an earlier date but later than mid-2001 when the photometric
observations by Henden \& Munari (2008) found it in quiescence.  The
amplitude of the outburst decreases with increasing wavelength (down to
$\Delta I_{\rm C}$$\sim$0.45 mag), as typical in symbiotic stars where the
cool giant is usually a passive bystander of the eruption.  From mid-2009 to
Jan 2016, StH$\alpha$ 169 has remained at flat quiescence, and when the
observations resumed in April 2016 we found the object declining from
maximum during a new outburst.  The start of the current outburst could be
marked by the last observation of the previous observing season, on 2016 Jan
21, when StH$\alpha$ 169 appears already brighter than any other previous
$B$-band quiescence observations (cf Figure~1).  The recorded peak
brightness for both the 2005 and 2016 outbursts is the same ($B$=13.6), as
it is the same the initial fast decline.  Only continued monitoring will
reveal if the current outburst will replicate the behaviour of the previous
eruption that was characterized by a much slower rate during the last
magnitude of decline.

\IBVSfig{9.5cm}{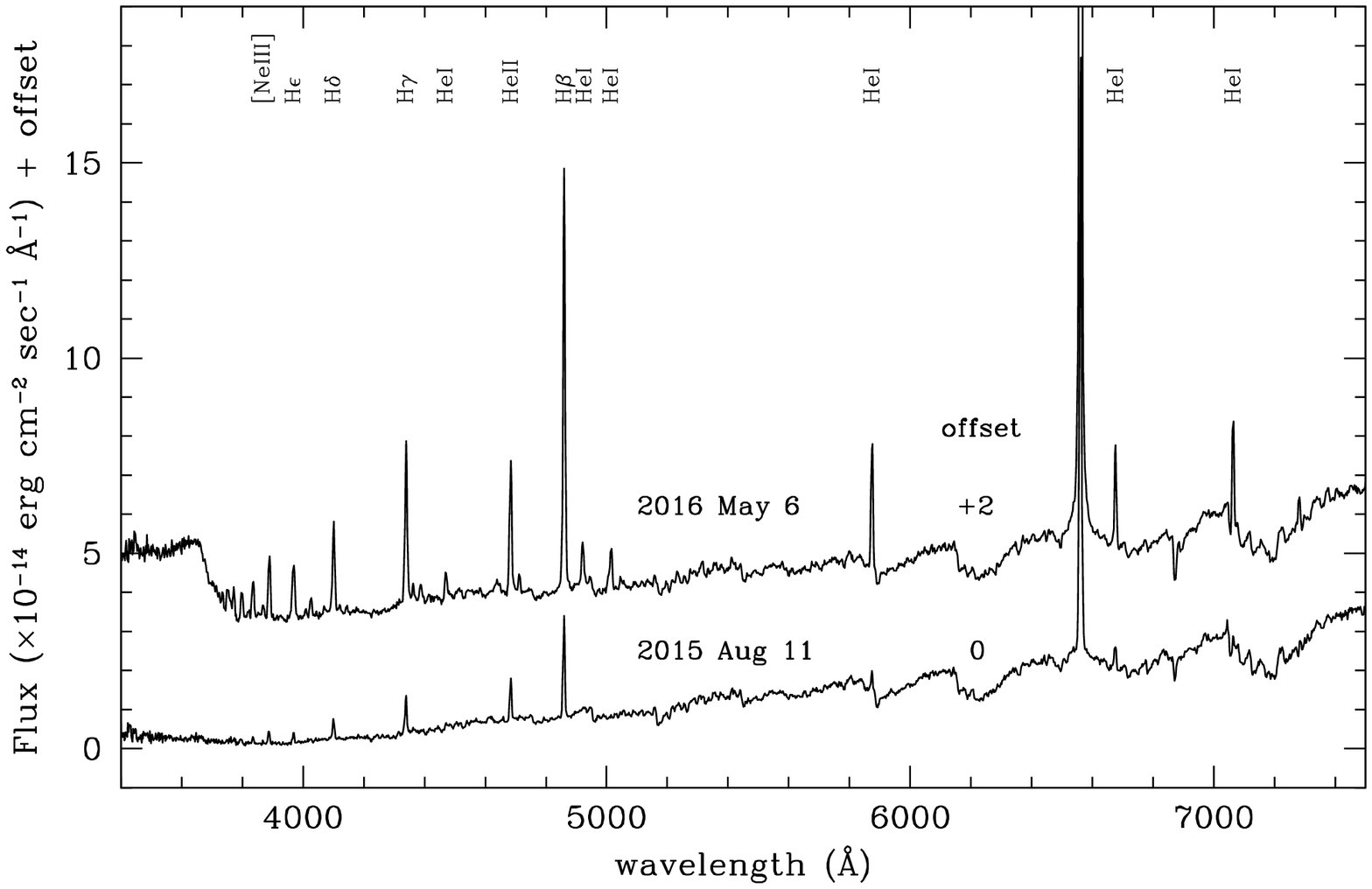}{Fluxed low resolution spectra of StH$\alpha$
169. The one for 2015 Aug 11 is typical of quiescence conditions, that for
2016 May 6 shows the appearance during the current outburst.}

Low resolution spectra of StH$\alpha$ 169 are regularly obtained with the
1.22m telescope + B\&C spectrograph operated in Asiago by the Department of
Physics and Astronomy of the University of Padova.  Figure~2 compares our
last spectrum of StH$\alpha$ 169 during the preceeding quiescence with one
obtained during the current outburst.  In outburst, a strong blue continuum
overwhelms the M giant absorption spectrum short of 5800 Ang, and the Balmer
continuum turns into strong emission.  The [NeV] 3426 and the OVI Raman
scattering at 6825 \AA, that are weakly present in quiescence, are gone. 
During outburst, the emission lines have largely increased their integrated
flux, Balmer lines by 7$\times$, HeII 4686 by 4.5$\times$ and HeI by
9$\times$.  The width of the emission lines remains sharp and the same as in
quiescence, and no P-Cyg profile is visible.

\IBVSfig{9.2cm}{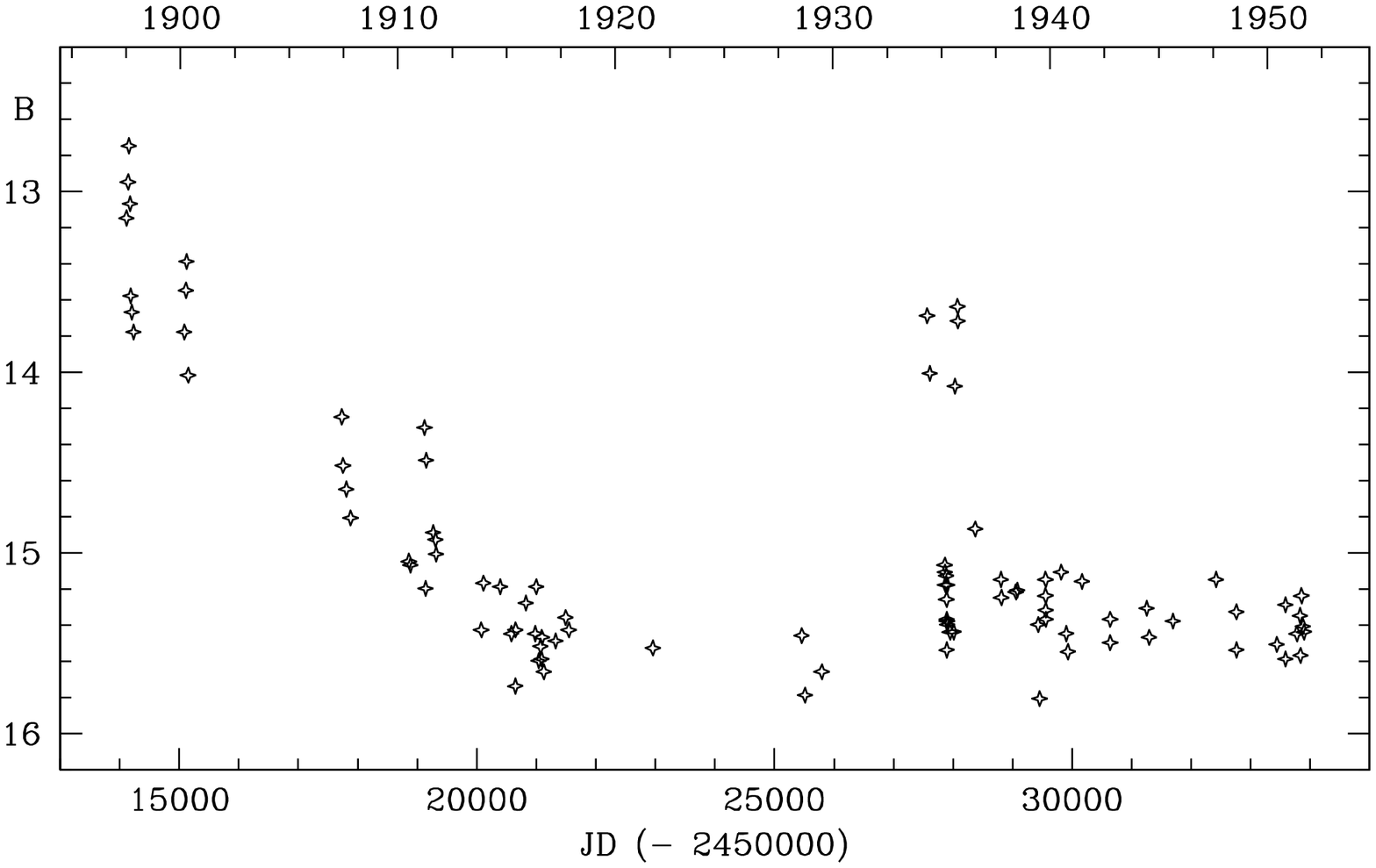}{$B$-band historical lightcurve of StH$\alpha$
169 from Harvard plates.}

To put our 2005-2016 CCD observations in a broader context, we have searched
via the DASCH database the Harvard plate archive for historical data on
StH$\alpha$ 169.  We found the star to have been positively recorded on 94
blue sensitive Harvard plates.  The corresponding lightcurve is plotted in
Figure~3, where the original DASCH data have been shifted by +0.14 mag to
match the modern $B$-band CCD scale.  This shift has been derived by
comparing the DASCH $B$-band magnitudes for the photometric comparison
sequence around StH$\alpha$ 169 with the values published by Henden and
Munari (2006).  The shifted DASCH magnitudes are listed in Table~2
(available electronic only).  The 1897-1951 lightcurve in Figure~3 is
characterized by a series of brightenings superimposed on a general declines
in brightness that affected StH$\alpha$ 169 untill 1916, when the star
settled on a quiescence characterized by the same mean $B$=15.29 value that we
measured for quiescence during 2009-2015.  Two rapidly evolving outburst
were recorded in 1934 and 1935, both peaking at $B$$\sim$13.7 about 510 days
apart.  Such peak brightness is remarkably similar to the $B$=13.6 value
characterizing both the 2005 and the 2016 events we have observed (cf
Figure~1).

We thank Alison Doane, Curator of Astronomical Photographs at the Harvard
College Observatory, for granting us access to DASCH database
(partially supported from NSF grants AST-0407380, AST-0909073, and
AST-1313370). We also acknowledge the assistance by S. Dallaporta, L.
Baldinelli and A. Maitan (ANS Collaboration) with some of the photometric
measurements reported in this paper.

\references 

Belczy{\'n}ski K., et al., 2000, A\&AS, 146, 407

Downes R.~A., Keyes C.~D., 1988, AJ, 96, 777

Henden A., Munari U., 2006, A\&A, 458, 339

Henden A., Munari U., 2008, BaltA, 17, 293

Li J., et al., 2015, RAA, 15, 1332   

Munari U., Moretti S., 2012, BaltA, 21, 22

Munari U., et al., 2012, BaltA, 21, 13

Munari U., Graziani M., 2016, ATel, 9036,

Pigulski A., Pojma{\'n}ski G., Pilecki B., Szczygie{\l} D.~M., 2009, AcA, 59, 33

Ramsay G., Hakala P., Howell S.~B., 2014, MNRAS, 442, 489

Stephenson C.~B., 1986, ApJ, 300, 779

\endreferences

\end{document}